\documentclass[pra,aps,showpacs,groupedaddress,superscriptaddress,twocolumn,longbibliography,notitlepage]{revtex4-1}

\usepackage[utf8x]{inputenc}
\usepackage{color}
\usepackage{bbm} 

\usepackage{amsfonts,amsmath,amssymb,stmaryrd}

\usepackage{graphicx}
\usepackage{subfigure}  % use for side-by-side figures
\usepackage{bbm} 
\usepackage{hyperref}
\usepackage[pdftex]{epsfig}
\usepackage{mathrsfs}
\usepackage{verbatim}
\usepackage{centernot}
\usepackage{ulem}

\renewcommand{\l}{\left(}
\renewcommand{\r}{\right)}

\newcommand{\hc}{\text{h.c.}}
\newcommand{\eff}{\text{eff}}
\newcommand{\qp}{\text{qp}}
\newcommand{\C}{\mathcal{C}}
\renewcommand{\L}{\mathcal{L}}
\newcommand{\TP}{{\rm TP}}
\newcommand{\I}{{\rm I}}

\newcommand{\bra}[1]{\langle#1|}
\newcommand{\ket}[1]{|#1\rangle}

\renewcommand{\H}{\hat{\mathcal{H}}}
\renewcommand{\c}{\hat{c}}

\newcommand{\cd}{\hat{c}^\dagger}

\newcommand{\bd}{\hat{b}^\dagger}
\renewcommand{\b}{\hat{b}}

\usepackage{array}

\usepackage{cancel,ifthen} \newcommand{\cmnt}[2][NoInPuT]{\ifthenelse{\equal{#1}{NoInPuT}}{}{{\color{red}\sout{#1}}} {\color{blue} #2}}

\usepackage{bm}	% bold in math mode
\renewcommand{\vec}[1]{\bm{#1}}

\begin{document}
\normalem	% changes \emph back to normal after introducing ulem package.

\title{Interferometric Measurements of Many-body Topological Invariants\\ using Mobile Impurities}

\author{F. Grusdt}
\email[Corresponding author email: ]{fgrusdt@physics.harvard.edu}
\affiliation{Department of Physics and Research Center OPTIMAS, University of Kaiserslautern, Germany}
\affiliation{Graduate School Materials Science in Mainz, Gottlieb-Daimler-Strasse 47, 67663 Kaiserslautern, Germany}
\affiliation{Department of Physics, Harvard University, Cambridge, Massachusetts 02138, USA}

\author{N. Y. Yao}
\affiliation{Department of Physics, University of California, Berkeley, California 94720, USA}

\author{D. Abanin}
\affiliation{Department of Theoretical Physics, University of Geneva, 24 quai Ernest-Ansermet, 1207 Geneva, Switzerland}

\author{M. Fleischhauer}
\affiliation{Department of Physics and Research Center OPTIMAS, University of Kaiserslautern, Germany}

\author{E. Demler}
\affiliation{Department of Physics, Harvard University, Cambridge, Massachusetts 02138, USA}


%\keywords{polaron transformation, topology, Mott insulator, Zak phase, Berry phase, coherent probe, impurity, Ramsey interferometry, Bloch oscillations, Chern number, Chern insulator, fractional quantum Hall effect, Laughlin states}

\date{\today}

\maketitle

\textbf{Topological quantum phases cannot be characterized by Ginzburg-Landau type order parameters, and are instead described by non-local topological invariants. Experimental platforms capable of realizing such exotic states now include ``synthetic" many-body systems such as ultracold atoms or photons. Unique tools available in these systems enable a new characterization of strongly correlated many-body states. Here we propose a general scheme for detecting topological order using interferometric measurements of elementary excitations. The key ingredient is the use of mobile impurities which bind to quasiparticles of a host many-body system. Specifically we show how fractional charges can be probed in the bulk of fractional quantum Hall systems. We demonstrate that combining Ramsey interference with Bloch oscillations can be used to measure Chern numbers of individual quasiparticles, which gives a direct probe of their fractional charges. We discuss possible extensions of our method to other topological many-body systems, such as spin liquids.}

Many-body systems with spontaneous symmetry breaking can be described by Ginzburg-Landau theories, formulated in terms of local order parameters. This powerful approach provides a universal description of systems with very different microscopic Hamiltonians but with similar type of symmetry breaking, such as superfluids and ferromagnets. The integer and fractional quantum Hall effects (IQHE and FQHE) \cite{Vonklitzing1980,Tsui1982,Laughlin1983} in contrast are examples of quantum phases of matter, for which no local order parameters exist. Instead, these systems are described by non-local topological invariants \cite{Wen1995}. The fractional charges of elementary excitations \cite{Tsui1982,Laughlin1983}, the many-body Chern number $\C$ \cite{Niu1985} and, in the case of quantum spin liquids, fractional quantum Hall systems and fractional Chern insulators \cite{Sorensen2005,Hafezi2007,Neupert2011FCI,Tang2011,Regnault2011,Sun2011,Sheng2011,Parameswaran2013}, the groundstate degeneracy on a torus \cite{Kitaev2006a}, constitute important examples of topological order parameters.

%%%%%%%%%%%%%%%%%%%%%%%%%%%%%%%%%%%%%%%%%%%%%%%%%%%%%
\begin{figure}[b]
\centering
\epsfig{file=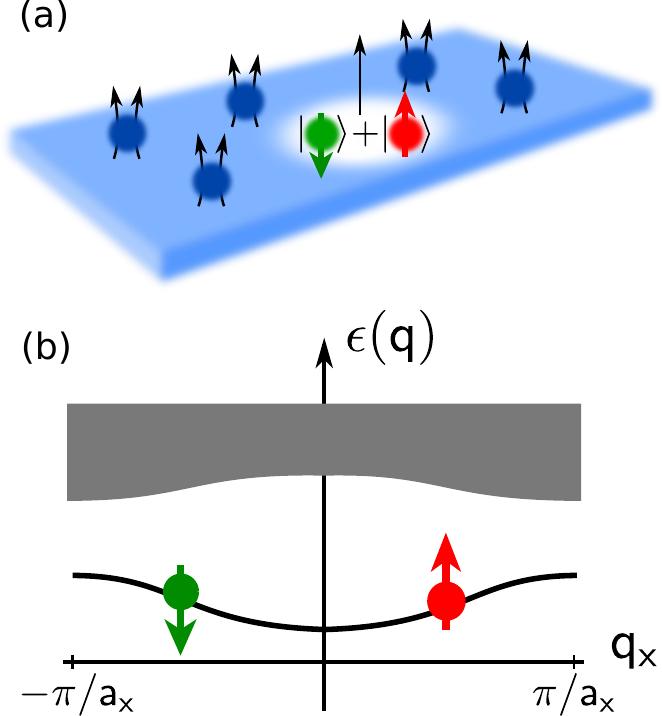, width=0.3\textwidth}
\caption{We propose a scheme for the measurement of many-body topological invariants of interacting states with topological order. It can be applied to measure the Chern numbers of Abelian quantum Hall states (for example of Laughlin type) and of their topological excitations, as illustrated in (a). An elementary excitation (here a quasihole) is coupled to a mobile two-component impurity. When the impurity is tightly bound to the quasihole, a topological polaron is formed. It can be labeled by its quasi-momentum $q$, and its two internal degrees of freedom $\uparrow,\downarrow$ and can be used to perform interferometry. The spectrum of the many-body system is depicted in (b) for a generic 1D case. The topological invariant of the qp band structure can be measured using tools developed for non-interacting systems by a combination of Bloch oscillations and Ramsey interferometry.}
\label{fig:intro}
\end{figure}
%%%%%%%%%%%%%%%%%%%%%%%%%%%%%%%%%%%%%%%%%%%%%%%%%%%%%

Probing -- and, in some cases, even defining -- the non-local order parameters of topological systems with strong correlations represents a major experimental and theoretical challenge. Many indicators used in the theoretical description of such systems, like the entanglement entropy \cite{Kitaev2006,Levin2006} and spectrum \cite{Li2008}, are difficult to probe directly in current experiments, although first steps in this direction have been undertaken \cite{Abanin2012Ent,Daley2012,Mazza2015,Islam2015}. Previously it has been shown that edge excitations can be used to detect topological orders by measuring their fractional charges \cite{Wen1995,Goldman1995,Picciotto1997,Saminadayar1997} and statistics \cite{Wen1995}. 

Here we introduce an alternative approach which allows one to measure topological order parameters directly in the bulk of the system, without the need of relying on the bulk-edge correspondence. The main idea is to map out the topology characterizing the effective bandstructure of elementary quasiparticle (qp) excitations. As will be pointed out, it is intimately related to the topological order of the groundstate. In particular, we show how the Chern numbers of the effective qp bandstructures can be measured by combining Bloch oscillations with a Ramsey interferometric sequence, see FIG.\ref{fig:intro}. We point out that they are directly related to the corresponding (fractional) charges for arbitrary Abelian quantum Hall states, and show how the Chern number of the many-body groundstate can be derived. Our scheme extends earlier ideas \cite{Atala2012,Abanin2012,Grusdt2014Z2,Duca2014}, which have been developed to measure topological invariants of essentially non-interacting particles (ultracold atoms in particular), to the realm of strongly correlated quantum many-body systems. More generally, our interferometric method paves the way for a detailed investigation of qp properties, including, possibly, their braiding statistics.

Our method is ideally suited to systems of ultracold atoms, which recently emerged as a new promising platform for realizing and probing various topological states of matter. The ability to perform interferometric measurements in such systems is one of their key technical advantages in comparison with other experimental setups. Cold atoms provide a versatile toolbox, allowing to engineer not only single-particle properties of Hamiltonians, like the shape of optical lattice potentials, but also the interactions between particles \cite{Bloch2008,Lahaye2009,Saffman2010,Chin2010}. Recently the Chern number has been measured in transport experiments \cite{Aidelsburger2014} and the celebrated Haldane model has been realized \cite{Jotzu2014} in systems of weakly interacting ultracold atoms. An experimental realization of the FQHE in such systems \cite{Wilkin1998,Regnault2003,Sorensen2005,Hafezi2007,Baranov2008,Grusdt2013c,Yao2013} should be within reach with the currently available tools. In addition, direct and fully coherent control over individual atoms has been demonstrated in experiments with ultracold quantum gases, see e.g. Refs. \cite{Frese2000,Karski2009}. In the present article we present a new concept of using impurity atoms as coherent probes of the topological invariants of strongly correlated many-body systems of host atoms. 

%%%%%%%%%%%%%%%%%%%%%%%%%%%%%%%%%%%%%%%%%%%%%%%%%%%%%
%%%%%%%%%%%%%%%%%%%%%%%%%%%%%%%%%%%%%%%%%%%%%%%%%%%%%
\section*{Results}
%%%%%%%%%%%%%%%%%%%%%%%%%%%%%%%%%%%%%%%%%%%%%%%%%%%%%

The key idea of our approach is to measure the Chern number of the effective qp bandstructure using a generalization of the interferometric technique developed for non-interacting systems in Refs. \cite{Atala2012,Abanin2012}. First, let us briefly summarize the main idea of the interferometric protocol for a weakly interacting Bose-Einstein condensate loaded in a two-dimensional Bloch band, in a system with an effective magnetic field. The first ingredient to measure the corresponding Chern number $\C$, is a direct detection of the geometric Zak phase $\varphi_{\rm Zak}(k_y)$ \cite{Atala2012} for a given value of the quasi-momentum $k_y$. 

To this end, the condensate is moved to $(k_x,k_y)$, where a $\pi/2$ Ramsey pulse is used to prepare a superposition of two internal states $\sigma=\uparrow,\downarrow$. The initial wavefunction thus reads $\l \ket{\uparrow}  + \ket{\downarrow} \r \otimes \ket{k_x,k_y}/ \sqrt{2}$. Next, sufficiently weak opposite forces $\vec{F}(\downarrow) = - \vec{F}(\uparrow)$ are applied such that the two components undergo Bloch oscillations, see FIG. \ref{fig:intro} (b). When the force is applied along the $x$-axis for a time $\Delta t = G_x / 2 F$, where $G_x$ denotes the width of the magnetic Brillouin zone (BZ) containing an integer number of magnetic flux quanta, the two components pick up a relative geometric phase $\varphi_{\rm Zak}(k_y)$ \cite{Zak1989} and the wavefunction reads $\l \ket{\uparrow}  + e^{i \varphi_\text{Zak}(k_y)} \ket{\downarrow} \r \otimes \ket{G_x/2,k_y}/ \sqrt{2}$. By recombining the two states using a second $\pi/2$ Ramsey pulse, the Zak phase can be read out. This measurement can be repeated for different values of $k_y$ \cite{Abanin2012}, and the winding of the Zak phase across the magnetic BZ in $k_y$-direction (size $G_y$) gives the Chern number \cite{Xiao2010},
\begin{equation}
\C = \frac{1}{2 \pi} \int_0^{G_y} dk_y ~ \partial_{k_y} \varphi_{\rm Zak}(k_y).
\end{equation}
For more details, including the discussion of dynamical phases and gauge dependence of the Zak phase, the reader is referred to Refs.\cite{Atala2012,Abanin2012,Grusdt2014Z2,Duca2014}.

Now we extend the interferometric protocol to strongly correlated systems. The key idea is to apply the exact same sequence as described above to a single qp excitation in the host system. To obtain fully coherent control of the qp, it is coupled to a mobile impurity. The resulting composite object (the impurity bound to the topological excitation) will be called a \emph{topological polaron} (TP), which is at the heart of our scheme. We assume that the impurity has an internal degree of freedom $\sigma$, which allows one to perform Ramsey interferometry as described above. By applying opposite forces $\sigma^z F \vec{e}_x$ directly to the two impurity components for the same time $\Delta t = G_x / 2 F$ as in the non-interacting case (where $G_x$ is defined for host particles), a topological invariant can be measured which will be identified as the Chern number $\C_\TP$ of the effective TP bandstructure (for more details see Methods Section). Note that now $\ket{k_x,k_y}$ stands for the full many-body wavefunction describing the TP at quasi-momentum $(k_x,k_y)$, and in this sense true many-body Zak phases $\varphi_{\rm Zak}(k_y)$ are measured. 

Now our main results are summarized. We will establish that the Chern number of the TP bandstructure is directly related to the fractional charges of the qps. We will use Chern-Simons effective field theory for the description of Abelian fractional quantum Hall states and their excitations. In addition we introduce a numerically exact technique for calculating Zak phases of TPs for small systems. As a concrete example we will show for Laughlin states at the filling fraction $\nu=1/m$ (defined as a ratio of particle density to flux density) that the TP Chern number, discussed in the protocol above, is given by the inverse of the fractional qp charge $e^*=e/m$ ($e$ is the charge of particles in the host many-body system), 
\begin{equation}
\C_\TP = \frac{e}{e^*} = m.
\label{eq:CTPrelationK}
\end{equation}
Therefore a measurement of the TP Chern number directly yields the fractional qp charge. The result in Eq.\eqref{eq:CTPrelationK} moreover indicates a direct relation to the many-body Chern number $\C_0=1/m$ \cite{Niu1985} of the incompressible $\nu=1/m$ Laughlin groundstate, $\C_0 = 1/\C_\TP$. Similar relations are true for any Abelian quantum Hall state, where $e/e^*$ needs to be expressed in terms of the $K$-matrix. In that case there exist $J=1,...,n$ different qp sectors, with fractional charges $e^*_J$ and qp Chern numbers $C_\TP^{(J)} = e/e^*_J$. The Chern number of the incompressible fractional quantum Hall state is given by $\C = \sum_{J=1}^n 1 / \C_\TP^{(J)}$, see Methods Section.

%%%%%%%%%%%%%%%%%%%%%%%%%%%%%%%%%%%%%%%%%%%%%%%%%%%%%%%
~ \\
\textbf{Topological polarons and qp Chern numbers}
\label{sec:Topolarons}
%%%%%%%%%%%%%%%%%%%%%%%%%%%%%%%%%%%%%%%%%%%%%%%%%%%%%%%
Our discussion of TPs will be organized as follows. In this section we focus on the physical picture of TPs in fractional quantum Hall systems. We discuss the origin of equation \eqref{eq:CTPrelationK} assuming that an impurity particle is strongly bound to a qp so the impurity current $j_\mu^\sigma$ (where $\mu=x,y,t$ and $\sigma=\uparrow, \downarrow$) is equivalent to the qp current. Detailed discussion of the binding mechanism between the impurity and a qp will be presented in the next subsection.

As in the theoretical proposal of Ref.\cite{Abanin2012} and the experimental demonstration of Ref.\cite{Atala2012} we assume that the two-component TP can be controlled by the force $\vec{F}$ acting on the impurity. (In experiments a spin-dependent force $\vec{F}$ could be realized by a magnetic field gradient.) We model this by coupling the TP current $j_\mu^\sigma$ to an external field that acts differently on the two internal states of the impurity but does not affect host particles. We describe this field by the potential $B_\mu(\sigma)$ so that the external force acting on the impurity is given by $F_i(\sigma) = - q \l \partial_t B_i + \partial_i B_0 \r$, where $q$ denotes the charge of the impurity associated with the field $B_\mu$. We emphasize that the impurity is not affected directly by the gauge field creating a fractional quantum Hall state for host particles. However by binding a qp of the surrounding fractional quantum Hall state it acquires topologically non-trivial dynamics. This can be understood as arising from the particle-vortex duality of the FQHE, which implies that the moving qp bound to the impurity ``sees" surrounding host particles as a source of effective flux. 

Using the Chern-Simons description of  the FQHE one can show (see Methods Section for details) that the effective Lagrangian describing the TP is given by (throughout the paper we set $\hbar = 1$)
\begin{equation}
\L_{\rm TP} = \sum_{\sigma=\uparrow, \downarrow} \l q B_\mu(\sigma) + e^* A_\mu \r j_\mu^\sigma.
\label{eq:LeffTP}
\end{equation}
Here $A_\mu$ denotes the external gauge field corresponding to the homogeneous magnetic field $b_z$ seen by the host particles in the many-body quantum Hall system, $\partial_x A_y - \partial_y A_x = b_z$. The second term in the Lagrangian \eqref{eq:LeffTP} describes the emergent coupling of the TP to the gauge field $A_\mu$, albeit with a fractional $A_\mu$-charge $e^* = e /m$. Hence according to Eq.\eqref{eq:LeffTP} the TP sees an effective magnetic field $b_z^* = b_z e^*/e$. 

We emphasize again that the emergent coupling of impurity particles to the gauge field $A_\mu$ arises only through binding of a qp excitation. For Laughlin states, a qp acquires an Aharonov-Bohm phase $2 \pi$ when going around a single host particle. Hence the effective ``flux density" seen by the qp is equal to the density of host particles, which is $1/m$ times the magnetic flux density of $A_\mu$.

Now the Chern number $\C_\TP$ of the TP can easily be calculated. To this end we note that, in a homogeneous magnetic field $b_0$, the Berry curvature $\mathcal{F}$ is constant, $\mathcal{F} =1 / (e b_0)$. The way we described the interferometric protocol above, we defined the TP Chern number by integrating the Berry curvature over the entire magnetic BZ of the host many-body system. Its size is $G_x \times G_y = 2 \pi e b_z$, where $b_z$ is the magnetic field seen by host particles. Because the Berry curvature seen by the TP is $\mathcal{F}_\TP =1 / (e b_z^*)$ we obtain
\begin{equation}
\C_\TP = \frac{1}{2\pi} \int_{\rm BZ} d^2k ~ \mathcal{F}_\TP = \frac{b_z}{b_z^*} = \frac{e}{e^*},
\label{eq:CTPderivation}
\end{equation}
as claimed in Eq.\eqref{eq:CTPrelationK}. 

Note that because of gauge invariance $\C_{\TP} \in \mathbb{Z}$ has to be an integer \cite{Thouless1982,KOHMOTO1985}, as long as the TP groundstate is non-degenerate for all $k$ (see also Methods Section). Therefore, when the qp charge is not a fraction of one, the band of TP groundstates needs to have degeneracies. When $e^*/e=p/q$, where $p$ and $q$ are relative prime, we expect $p$ degenerate bands sharing a total Chern number of $\C^{\rm tot}_\TP=q$. We find numerical signatures for this degeneracy, and a specific example is discussed in the supplementary.

Alternatively, the qp charge $e^*/e$ of the TP could be obtained directly from an interferometric measurement of the Aharonov Bohm phase $\Phi_{\rm AB} = 2 \pi \frac{e^*}{e} A b_z/\Phi_0$ in the magnetic field $b_z$. Here $A$ denotes the area encircled by the TP and $\Phi_0$ is the magnetic flux quantum.  Because obtaining control over the impurity momentum is technically less challenging than achieving full spatial resolution, the interferometric measurement of the Chern number should be easier to implement.

%%%%%%%%%%%%%%%%%%%%%%%%%%%%%%%%%%%%%%%%%%%%%%%%%%%%%%%
~ \\
\textbf{Microscopic description}
\label{sec:MicroModel}
%%%%%%%%%%%%%%%%%%%%%%%%%%%%%%%%%%%%%%%%%%%%%%%%%%%%%%%
Now we provide a microscopic description of TPs in correlated many-body systems interacting with a single impurity particle. To this end we investigate concrete models of interacting fermions on a lattice, whose groundstates are integer and fractional Chern insulators (ICI and FCI respectively) \cite{Sorensen2005,Hafezi2007,Neupert2011FCI,Tang2011,Regnault2011,Sun2011,Sheng2011}. We develop an exact numerical method to solve this problem and compare our results to an approximate strong coupling theory of TPs which we also introduce below. Like before, we concentrate on states in the Laughlin universality class, with a single qp sector.

We consider a topologically ordered phase in a many-body system (referred to as the host system), described by the groundstate of a Hamiltonian $\H_0$. For probing qp excitations, we introduce a mobile impurity with two internal states, described by $\H_\I =  \H_\I^0 - \hat{\sigma}^z  \vec{F} \cdot \hat{\vec{r}}$. Here $\H_\I^0$ denotes the kinetic energy and $\hat{\vec{r}}$ the position operator of the impurity. The two internal states $\uparrow$ and $\downarrow$ of the impurity experience opposite forces $\pm \vec{F}$, as described by the Pauli matrix $\hat{\sigma}^z$. To bind qps to the impurity, a local interaction $\H_\text{int}$ with the many-body system is introduced. Its concrete form can differ from model to model, but for simplicity we will assume throughout that it is independent of the internal state of the impurity. Thus our system is described by the Hamiltonian $\H = \H_0 + \H_\I + \H_\text{int}$. Local interactions between particles with multiple internal states are routinely realized in systems of ultracold atoms.

In equilibrium, i.e. for $\vec{F}=0$, the groundstate $\ket{\psi_\TP(\vec{q},\sigma)}$ describes a TP and can be labeled by its quasi-momentum $\vec{q}$ and its internal state $\sigma = \uparrow,\downarrow$. The external force $\vec{F}$ couples to the quasi-momentum $\vec{q}$ of the TP. This drives Bloch oscillations where the quasi-momentum changes according to $\frac{d}{dt} \vec{q} = \sigma^z \vec{F}$, where $\sigma^z=1$ ($\sigma^z=-1$) for $\sigma=\uparrow$ ($\sigma=\downarrow$), respectively). By applying the scheme described in Ref. \cite{Atala2012} to the states $\ket{\psi_\TP(\vec{q},\sigma)}$ the geometric Zak phases characterizing the TP band structure can be measured, see FIG.\ref{fig:intro}. As discussed in the Methods Section, the corresponding Chern number is obtained by integrating the Berry curvature seen by the TP over the magnetic BZ of the host many-body system.

\emph{Strong coupling theory.--}
Before turning our attention to concrete models, we introduce the strong coupling theory of TPs which is inspired by Landau's and Pekar's treatment of the polaron problem in polar crystals \cite{Landau1946,Landau1948}. There are several important energy scales for describing TPs. Firstly qp excitations are characterized by the bandwidth $J_\qp$ of their effective dispersion and by the energy required for their creation which is set by the bulk excitation gap $\Delta_0$. Secondly the impurity is characterized by the effective hopping $J_\I$ and the coupling strength to the host particles $V$. Our description of TPs requires the following hierarchy of scales. The qp gap $\Delta_0$, which is larger than $J_\qp$, should be larger than $J_\I$ and $V$, i.e. $\Delta_0 \gg J_\I, V $. The impurity-host particle interaction strength $V$ should be chosen such that the impurity binds precisely one qp. We also need $J_\qp$ to be smaller than $J_\I$, i.e. $J_\qp \ll J_\I$, so that the total momentum of the TP, which is effectively shared by the impurity and the qp, resides predominantly in the qp. Another way of understanding this requirement is that the impurity should be fast compared to the qp and thus follow its dynamics  adiabatically.

Next we introduce a frame where the total momentum of the TP is conserved explicitly. To this end we restrict ourselves to a single qp and approximate $\H_0 \approx \sum_{\vec{k}}  \epsilon_\qp(\vec{k}) ~\ket{\psi_\qp(\vec{k})} \bra{\psi_\qp(\vec{k})}$, with $\epsilon_\qp(\vec{k})$ being the effective dispersion of qps and $\ket{\psi_\qp(\vec{k})}$ the qp state with momentum $\vec{k}$. We apply the unitary transformation $\hat{U}_{\rm LLP} = e^{i \hat{\vec{p}} \cdot \hat{\vec{R}}_\qp}$ introduced by Lee, Low and Pines (LLP) \cite{Lee1953}, where $\hat{\vec{p}}$ is the impurity momentum operator and $\hat{\vec{R}}_\qp$ is qp position operator conjugate to its momentum operator $\hat{\vec{P}}_\qp = \sum_{\vec{k}} \vec{k} \ket{\psi_\qp(\vec{k})} \bra{\psi_\qp(\vec{k})}$.  The transformed Hamiltonian $\hat{U}_{\rm LLP}^\dagger \H \hat{U}_{\rm LLP}$ reads
\begin{multline}
\H_{\rm SC} =  \H_\I^0 + \H_{\rm int}\bigl( \vec{R}_\qp=\vec{0} \bigr)  - \sigma^z \vec{F} \cdot \bigl( \hat{\vec{r}} + \hat{\vec{R}}_\qp \bigr) + \\ 
+ \sum_{\vec{k}}  \epsilon_\qp(\vec{k} - \hat{\vec{p}}) ~\ket{\psi_\qp(\vec{k})} \bra{\psi_\qp(\vec{k})},
\label{eq:HLLPstrongCoupling}
\end{multline}
as will be explained now.

The kinetic part of the impurity Hamiltonian commutes with the impurity momentum, $[\hat{\vec{p}},\H_\I^0]=0$, and remains unchanged. $\H_{\rm int}\bigl( \vec{R}_\qp=\vec{0} \bigr)$ denotes the interaction Hamiltonian for a qp localized in the origin of the new polaron frame. Here we assumed that (within the single qp approximation) $\H_{\rm int}=\H_{\rm int}(\hat{\vec{r}} - \hat{\vec{R}}_\qp)$ depends only on the relative distance between impurity and qp, and we used that $\hat{U}_{\rm LLP}^\dagger \hat{\vec{r}} \hat{U}_{\rm LLP} = \hat{\vec{r}} + \hat{\vec{R}}_\qp$. Finally because $[\hat{\vec{R}}_\qp,\hat{\vec{P}}_\qp]=i$ the qp momentum is shifted by an amount $\hat{\vec{p}}$ in the last line, $\vec{k} \to \vec{k} - \hat{\vec{p}}$.

Under the strong coupling conditions outlined above we can make a product ansatz for the TP wavefunction, $\ket{\psi_\TP(\vec{q})} = \ket{\psi_\qp(\vec{q})} \otimes \ket{\phi_\I}$, where the impurity follows the qp adiabatically. In the polaron frame the impurity sees a quasi-static potential created by the qp and its wavefunction $\ket{\phi_\I}$ is determined by the strong coupling impurity Hamiltonian
\begin{equation}
\H_{\I,\rm SC}  = \H_\I^0  - \sigma^z \vec{F} \cdot \hat{\vec{r}} + \H_{\rm int}\bigl( \vec{R}_\qp=\vec{0} \bigr). 
\label{eq:SCimpurityHamiltonian}
\end{equation}
This leads to a modification of the effective qp dispersion in Eq.\eqref{eq:HLLPstrongCoupling}, which we approximate by $\bra{\phi_\I} \epsilon_\qp(\vec{k} - \hat{\vec{p}}) \ket{\phi_\I}$. 
We proceed by eliminating the last term in the first line of Eq.\eqref{eq:HLLPstrongCoupling} by a time-dependent gauge transformation $\hat{U}_\qp(t) = e^{i \hat{\vec{R}}_\qp \cdot \hat{\sigma}^z \vec{F} t}$. In the resulting effective Hamiltonian the force $\vec{F}$ couples to the conserved momentum $\vec{k}$ and the qp Hamiltonian in strong coupling theory thus reads
\begin{equation}
\H_{\qp,\rm SC} =  \sum_{\vec{k}} ~  \bra{\phi_\I} \epsilon_\qp(\vec{k} - \hat{\sigma}^z \vec{F} t - \hat{\vec{p}}) \ket{\phi_\I} ~ \ket{\psi_\qp(\vec{k})}\bra{\psi_\qp(\vec{k})}.
\label{eq:HTPgeneric}
\end{equation}

%%%%%%%%%%%%%%%%%%%%%%%%%%%%%%%%%%%%%%%%%%%%%%%%%%%%%
\begin{figure*}[t]
\centering
\epsfig{file=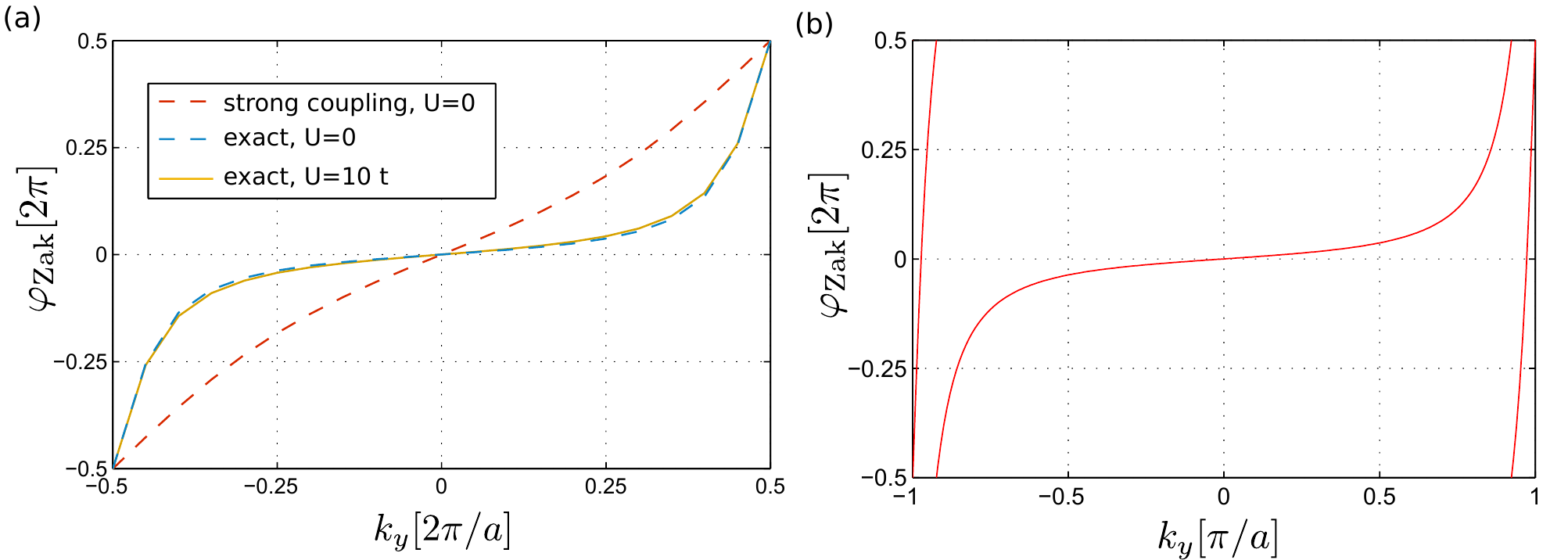, width=0.8\textwidth}
\caption{A single mobile impurity can be coupled to an elementary hole excitation of an integer (a) or fractional (b) Chern insulator, and form a TP. The winding of the many-body Zak phase $\varphi_{\rm Zak}(k_y)$ across the BZ defines the many-body Chern number of the TP. In (a) we compare predictions for ICIs without ($U=0$) and with inter-fermion interactions ($U \neq 0$). Parameters are $J=t/2$, $V=2t $, $\alpha=1/4$ and we simulated $4 \times 4$ sites with $N=3$ fermions. In (b) a $\nu=1/3$ FCI is considered, for parameters $J=t/2$, $V=2t $, $\alpha=1/4$, $U=10 t$. We simulated $4 \times 7$ sites filled with $N=2$ fermions, corresponding to $3N+1=7$ flux quanta as required for having one quasihole excitation.}
\label{fig:ICIresults}
\end{figure*}
%%%%%%%%%%%%%%%%%%%%%%%%%%%%%%%%%%%%%%%%%%%%%%%%%%%%%

Next we discuss how the topological invariant measured by the TP relates to the Chern number of the effective qp bandstructure using the strong coupling wavefunction. During the protocol described above (see also FIG.\ref{fig:intro}) the force $\hat{\sigma}^z \vec{F}$ is applied and the TP wavefunction $\ket{\psi_\TP(\vec{q},t)} = e^{i \varphi} \ket{\psi_\qp(\vec{q} - \sigma^z \vec{F} t)} \otimes \ket{\phi_\I}$ follows its groundstate adiabatically. Thereby it picks up a phase $\varphi$ containing a geometric part of $2 \pi \nu_\TP$ which is measured by the Ramsey sequence (dynamical phases are discussed in Refs. \cite{Atala2012,Abanin2012}). Due to the product form of the strong coupling wavefunction we find two contributions, $\nu_\TP = \nu_{\qp} + \nu_{\I}$. 

The first contribution is picked up by the qp wavefunction, $2 \pi \nu_{\qp} =\oint d\vec{k} \cdot \bra{\psi_\qp(\vec{k})} i \vec{\nabla}_{\vec{k}} \ket{\psi_\qp(\vec{k})}$. When the path in momentum space described by the TP in the interferometer encloses the (magnetic) BZ, the qp invariant is related to the TP Chern number defined above, 
\begin{equation}
\nu_{\qp} = \C_\TP.
\end{equation}
The second contribution $\nu_\I$ is picked up by the impurity part of the wavefunction. In the adiabatic limit of small $\vec{F}$ it is $2 \pi \nu_\I =- \sigma^z  \oint dt \vec{F} \cdot \bra{\phi_\I} \hat{\vec{r}} \ket{\phi_\I}$. This term corresponds to a geometric phase because it does not vanish in the limit when $\vec{F} \to 0$ and needs to be considered in general. It measures the displacement of the impurity wavefunction relative to the qp located in the origin of the polaron frame (recall that $\vec{R}_{\qp} = \vec{0}$ in the interaction Hamiltonian) and can become relevant in lattice systems. However when a closed loop $\oint dt \vec{F} =0$ is considered as in the interferometric sequence we discuss, the impurity invariant vanishes, $\nu_\I = 0$.

\emph{Topological polarons in Chern insulators.--}
Now we turn our attention to a concrete model of interacting particles on a lattice, described by the Hofstadter-Hubbard Hamiltonian
\begin{multline}
\H_0 =  - t \sum_{m,n} \left[ e^{- i 2 \pi \alpha n} \cd_{m+1,n} \c_{m,n} + \cd_{m,n+1} \c_{m,n} + \hc \right]  +\\ 
+ U \sum_{\langle (m,n), (m',n')\rangle} \cd_{m,n} \c_{m,n}  \cd_{m',n'} \c_{m',n'}.
\label{eq:H0fermiHofstadter}
\end{multline}
The first term is the celebrated Hofstadter model \cite{HOFSTADTER1976} and it describes free particles hopping between the sites $(m,n)$ of a square lattice in a magnetic field (using Landau gauge), where $\alpha$ denotes the magnetic flux density per plaquette (in units of the flux quantum) and $t$ is the hopping amplitude. The second term describes nearest neighbor interactions of strength $U$ between the particles. Here we consider fermions for concreteness, $\{ \c_{m,n}, \cd_{m',n'} \} = \delta_{m,m'} \delta_{n,n'}$, but a similar Hofstadter-Hubbard model has also been discussed for bosons with contact interactions \cite{Sorensen2005,Hafezi2007}. For sufficiently small values of $\alpha$, the groundstates of \eqref{eq:H0fermiHofstadter} show the IQHE and FQHE depending on the filling fraction $\nu$.

To study TPs we consider the Hofstadter-Hubbard model \eqref{eq:H0fermiHofstadter} at filling $\nu=1/m$ on a torus. We choose the number of flux quanta in the host many-body system $N_\phi=N m +1$ such that the groundstate of $\H_0$ contains one quasihole excitation. Next we add a single impurity, described by $\bd_{m,n}$, hopping between the sites of the same two-dimensional lattice,
\begin{multline}
\H_\I =- J\sum_{m,n} \left[ \bd_{m+1,n} \b_{m,n} + \bd_{m,n+1} \b_{m,n} + \hc \right]  - \\
- \vec{F} \cdot \sum_{m,n} \vec{r}_{m,n} ~ \bd_{m,n} \b_{m,n}.
\label{eq:HimpTPhofstadterICI}
\end{multline}
To bind the impurity to the quasihole, its interaction with the surrounding fermions is modeled by a repulsive contact potential, $\H_\text{int} = V \sum_{m,n} \cd_{m,n} \c_{m,n} \bd_{m,n} \b_{m,n}$. In the following we consider an impurity with only a single internal state for simplicity.

The model proposed above, Eqs.\eqref{eq:H0fermiHofstadter}, \eqref{eq:HimpTPhofstadterICI}, can be implemented with ultracold atoms. In Refs \cite{Aidelsburger2013,Miyake2013} the Hofstadter Hamiltonian was realized for bosons, and interactions can be introduced by using deep optical lattices. The impurity could be realized by adding a second atomic species, with different meta-stable internal states.

%%%%%%%%%%%%%%%%%%%%%%%%%%%%%%%%%%%%%%%%%%%%%%%%%%%%%
\begin{figure*}[t]
\centering
\epsfig{file=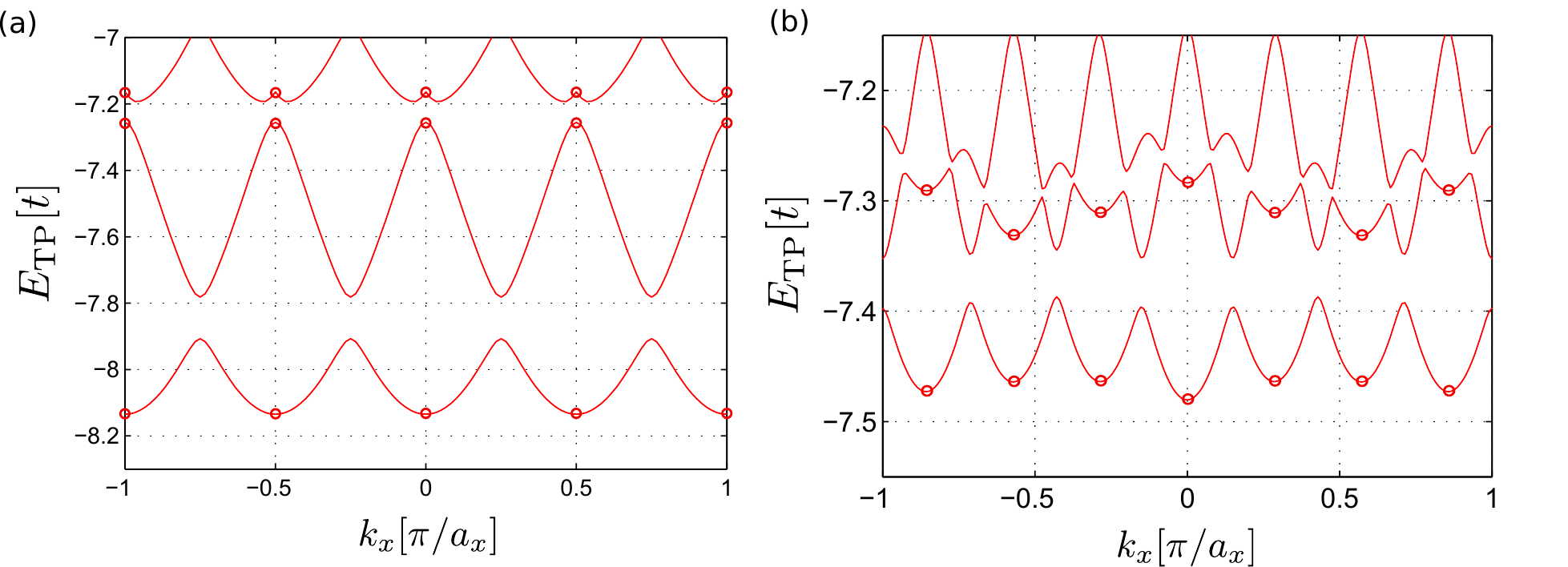, width=0.8\textwidth}
\caption{The energies of the lowest TP eigenstates are shown for $k_y=0$ as a function of the lattice moment $k_x$, calculated in the polaron frame (see Methods Section). In (a) an integer Chern insulator is considered and $U=10 t$ was used. Other parameters are the same as in FIG.\ref{fig:ICIresults} (a). In (b) the case of a fractional Chern insulator is shown, for parameters as in FIG.\ref{fig:ICIresults} (b). Circles mark the positions of the discrete momenta defined by the finite quantization volume.}
\label{fig:FCIresults}
\end{figure*}
%%%%%%%%%%%%%%%%%%%%%%%%%%%%%%%%%%%%%%%%%%%%%%%%%%%%%

In the Methods Section we present a formalism for calculating the full many-body TP wavefunction $\ket{\psi_\TP(\vec{q})}$ exactly for a given total momentum $\vec{q}$, based on the LLP polaron transformation. Here we use this approach to obtain both the dispersion relation and the Chern number of TPs. 

In FIG.\ref{fig:ICIresults} (a) our results are shown for an ICI, with and without fermion-fermion interactions $U$. We consider the case $\alpha=1/4$ and use Landau gauge where the size of the magnetic unit cell is $(a_x=4 a) \times a$, with $a$ denoting the lattice constant. We calculate the Chern number of the TP from the winding of the many-body Zak phase, $\C = \frac{1}{2 \pi} \int_{-\pi/a}^{\pi/a} dk_y ~ \partial_{k_y} \varphi_{\rm Zak}(k_y)$ \cite{Xiao2010}, see Methods Section for details. For comparison, the result of a simple strong coupling analysis is shown, where the impurity is bound to a free hole. Although the distribution of Berry curvature differs, the strong coupling theory predicts correctly the TP Chern number, $\C=1$. 

In FIG.\ref{fig:ICIresults} (b) we repeat our calculations for a FCI at $\nu=1/3$, corresponding to a $1/3$ Laughlin state. (We checked numerically that the incompressible groundstate of $\H_0$ has a fractional Chern number $\C_0=1/3$ and the expected three-fold groundstate degeneracy on a torus.) The Chern number of the TP, which is determined as the winding of the Zak phase of the TP in FIG.\ref{fig:ICIresults} (b), has the value $\C=3$. This is expected from the field-theoretical arguments given in the beginning, see Eq.\eqref{eq:CTPrelationK}.

An important quantity for estimating the robustness of our protocol is the energy gap of the TP state. To get some analytical understanding, we first consider the continuum limit $\alpha \to 0$ when the system enters the quantum Hall regime. Here the impurity (mass $M_{\rm I}$) interacts with fermions (mass $M_{\rm F}$) in a filling-$\nu$ Laughlin state through a local contact interaction $g_{\rm IF} \delta^{(2)}(\vec{r})$. Within the strong coupling theory introduced above, we find a finite TP binding energy given by
\begin{equation}
\Delta_{\rm TP} = \sqrt{\nu \frac{g_{\rm IF}}{2 \pi \ell_{\rm B}^2} \frac{M_{\rm F}}{M_{\rm I}} \omega_{\rm c} },
\end{equation}
where $\ell_{\rm B} = 1 /\sqrt{e b_0}$ is the magnetic length and $\omega_{\rm c}= 1 / (M_{\rm F} \ell_{\rm B}^2) $ the cyclotron frequency.

In FIG.\ref{fig:FCIresults} we calculate the TP spectra numerically for small lattice systems and find low-energy states where the impurity is bound to the quasihole forming the TP. We observe that the TP gap $\Delta_\TP$ shows large finite-size effects. The TP binding energy oscillates at a frequency given by the system size, but at the discrete momenta allowed by the finite quantization volume -- which are indicated in the figures by circles -- we find the largest values of the TP gap. In the case of ICI this TP gap constitutes a sizable fraction of the cyclotron gap $\Delta_{\rm TP} \lessapprox \omega_{\rm c} \approx 4 \pi \alpha t$. Note that $\omega_{\rm c}$ provides an upper bound for the TP gap, $\Delta_{\rm TP} < \omega_{\rm c}$, because at this energy unbound excitons can be created in the bulk. In the case of FCI in contrast, the TP gap is only a small fraction of the cyclotron energy, but so is the bulk gap $\Delta_{\rm LN}$ of the FCI, $\Delta_{\rm LN} < \omega_{\rm c}$ \cite{Hafezi2007}. In this case $\Delta_{\rm LN} $ similarly provides an upper bound for the TP gap as the cyclotron energy in the ICI case, $\Delta_{\TP} < \Delta_{\rm LN}$.

To form a TP for mapping out the topological order of the host many-body system, we wish to realize the strong coupling limit described above. Thus, for impurity dynamics to be fast compared to qp tunneling, a weak qp dispersion is desirable. This can be achieved in lattice systems by including long-range hoppings for the host particles, see Ref.\cite{Kapit2010}. We simulated TPs in systems with long-range tunnelings of this kind (see also supplementary) and verified that the TP Chern number $\C=3$ is a robust feature of the groundstates.

%%%%%%%%%%%%%%%%%%%%%%%%%%%%%%%%%%%%%%%%%%%%%%%%%%%%%%%
%%%%%%%%%%%%%%%%%%%%%%%%%%%%%%%%%%%%%%%%%%%%%%%%%%%%%%%
\section*{Discussion}
%%%%%%%%%%%%%%%%%%%%%%%%%%%%%%%%%%%%%%%%%%%%%%%%%%%%%%%
By coupling a mobile impurity to the topological qp excitations of an incompressible many-body groundstate of host particles, a TP can be formed. Using internal degrees of freedom of the impurity, fully coherent control can be gained over individual qp excitations of the host many-body system. We demonstrated that TPs can be used to measure the topological invariants characterizing such qp excitations. In particular we developed an interferometric measurement scheme for qp Chern numbers and showed that these are directly related to the fractional charges of topological excitations in two dimensions. To this end we generalized schemes developed earlier for non-interacting systems and showed by explicit calculations that our scheme can be applied to integer- and fractional quantum Hall systems and Chern insulators.

In systems of ultracold atoms, impurities can be realized e.g. by introducing another atomic species and their internal degree of freedom corresponds to different hyperfine states. To neglect the effects of interactions between TPs, the density of impurity atoms needs to be well below one atom per cyclotron orbit. We expect that, in order to realize TPs in an experiment, a main challenge will be their preparation. Because TPs can carry fractional charges as in the case of FCIs, their preparation requires non-local operations in general. Therefore we suggest to build impurity atoms into the state as defects already before preparing the incompressible many-body state of host atoms. One concrete approach would be to start from this system and cool it down into its groundstate. If the densities of majority and impurity atoms are suitably chosen, this leads to the formation of TPs. Alternative approaches have also been discussed in the literature \cite{Paredes2001,Grusdt2014LN,Zhang2014,Zhang2015}.

The method presented in this article can provide a powerful experimental tool for detecting topological order in interacting many-body systems. It is ideally suited for cold atom experiments, which offer a rich toolbox with precise and fully coherent control over individual atoms. The concept of topological polarons as experimental probes for topological order can be generalized to other systems however, including for example qp excitations on topological superconductors, or systems with symmetry-protected topological orders. Another interesting direction would be to probe the (non-Abelian) braiding statistics of anyons by coupling them to impurities and forming topological polarons. In this case, one can envision interferometric sequences designed to probe braiding statistics, which work in real-space rather than momentum-space, as considered in this work.

We expect that our idea -- which is to witness topological order in a groundstate by probing the topology characterizing the effective bandstructure of its qp excitations -- may also be of broader theoretical interest. It provides a direct route how the concepts developed for non-interacting systems can be generalized to correlated many-body systems: We can define a set of topological invariants for a many-body state by considering single-particle topological invariants of all quasiparticle excitations. These may include Chern numbers, $\mathbb{Z}_2$ invariants or quantized Zak phases. The approach may be useful, for example, in the study of fractional topological insulators \cite{Levin2009} or quantum spin liquid states \cite{Kitaev2006a}.

%%%%%%%%%%%%%%%%%%%%%%%%%%%%%%%%%%%%%%%%%%%%%%%%%%%%%%%
%%%%%%%%%%%%%%%%%%%%%%%%%%%%%%%%%%%%%%%%%%%%%%%%%%%%%%%
\section*{Methods}
%%%%%%%%%%%%%%%%%%%%%%%%%%%%%%%%%%%%%%%%%%%%%%%%%%%%%%%
\textbf{Chern number of topological polarons}
In the main text we generalized the interferometric protocol developed for the measurement of Chern numbers of non-interacting particles \cite{Atala2012,Abanin2012} to a single qp excitation in a strongly-correlated many-body system. The coupling to an impurity particle was necessary for adapting the interferometric protocol. In this way a topological invariant of the TP was defined which, as we will now argue, is the Chern number of the TP. 

Consider a qp state $\ket{\psi_\TP(k_x,k_y)}$ which is characterized by its quasi-momentum $(k_x,k_y)$. Let us assume that the state is non-degenerate for all quasi-momenta. The corresponding Chern number $\C_\TP$ is then defined by the quantized Hall response to an external force $\vec{F}$. In the case of TPs the external force $\vec{F}$ acts directly on the impurity and couples to the quasi-momentum according to $\frac{d}{dt} \vec{k} = \vec{F}$. Using the Kubo-formula, Thouless et al. \cite{Thouless1982} have shown that the Chern number can be defined as an integral of the TP Berry curvature $\mathcal{F}_\TP$ over the (magnetic) Brillouin zone (BZ),
\begin{equation}
\C_\TP = \frac{1}{2 \pi} \int_{\rm BZ} d^2k ~ \mathcal{F}_\TP(\vec{k}).
\end{equation} 
The Berry curvature $\mathcal{F}_\TP = \nabla_{\vec{k}} \times \bra{u_{\vec{k}}} i \nabla_{\vec{k}} \ket{u_{\vec{k}}}$ is defined through the Bloch wavefunction $\ket{u_{\vec{k}}}$ constructed from $\ket{\psi_\TP(\vec{k})}$. The magnetic BZ is defined by the periodicity of the qp Hamiltonian, including gauge-dependent vector potentials \cite{Zak1964}. The periodicity of the TP wavefunction $\ket{\psi_\TP(\vec{k} + \vec{G})}=\ket{\psi_\TP(\vec{k})}$, where $\vec{G}$ is a reciprocal lattice vector, guarantees the integer quantization of the Chern number \cite{Thouless1982,KOHMOTO1985}.

The magnetic BZ of the TP is determined by the microscopic details of the model for both the impurity as well as the host many-body system. Let us assume that the impurity either lives in the continuum (such that effectively the impurity lattice constant $a \to 0$ vanishes) or that the unit-cell of the impurity lattice fits into the magnetic unit-cell of the host many-body system an integer number of times in a commensurable way. In either case, the magnetic BZ of the TP is then equal to the magnetic BZ of the host many-body system. This explains why, in our interferometric protocol, the TP Chern number needs to be defined as the winding of the Zak phase of the TP $\varphi_{\rm Zak}(k_y)$ over the magnetic BZ of the host many-body system, $k_y \to k_y + G_y$.\\

\noindent
\textbf{Field theory of topological polarons}
Here we discuss a field theory description of TPs, allowing us to derive the topological invariants characterizing their effective bandstructure. We consider a more general situation than discussed in the main text and allow quantum Hall states with arbitrary Abelian topological order. The generalization of our interferometric protocol to this case is discussed in the end.

Our starting point is a field-theoretical description of the topologically ordered host many-body system. We consider an incompressible groundstate with Abelian topological order, described by a Chern-Simons theory of level $n$. Such theories are believed to classify all Abelian topological orders and are relevant e.g. for the hierarchical description of the FQHE \cite{Halperin1984} or multi-layer fractional quantum Hall systems. They are is characterized by the symmetric integer $n$-by-$n$ matrix $\uuline{K}$ and the charge vector $\uline{t}$. The Lagrangian is \cite{Wen1995}
\begin{multline}
\L = \frac{1}{4 \pi} \uline{a}^T_\mu \uuline{K} \partial_\nu \uline{a}_\lambda \epsilon^{\mu \nu \lambda} - \frac{e}{2 \pi} A_\mu \uline{t}^T \partial_\nu \uline{a}_\lambda \epsilon^{\mu \nu \lambda} + \\ + \sum_{I=1}^n a_{I \mu} \ell_I j_{I \mu} + \rm kin.~energy.
\label{eq:LCS}
\end{multline}
Here $\uline{a}_\mu = a_{I=1...n, \mu}$ are the auxiliary compact $U(1)$ gauge fields from which the conserved current $J_\mu$ of the many-body system can be derived, $J_\mu = \frac{e}{2 \pi} \sum_I \partial_\nu a_{I \lambda} \epsilon^{\mu \nu \lambda}$. As usual, $\mu,\nu,...=t,x,y$ denote temporal and spatial coordinates. 

The first two terms of the Lagrangian \eqref{eq:LCS} describe the response of the many-body system to the external $U(1)$ gauge field $A_\mu$. Here $e$ denotes the $A_\mu$-charge of the indistinguishable host particles constituting the many-body system. From the Euler-Lagrange equations the quantized Hall response is obtained, $J_\mu = \C \frac{e^2}{2 \pi} \epsilon^{\mu \nu \lambda} \partial_\nu A_\lambda$, where the many-body Chern number is given by\begin{equation}
\C =  \sum_{I, J=1}^n t_I \l K^{-1} \r_{I J}.
\label{eq:manyBodyChernKtheory}
\end{equation}
The third term in \eqref{eq:LCS} describes the conserved currents $j_{I \mu}$ of the $I=1...n$ different qps (in the bulk of the system). The integers $\ell_I$ denote the number of qps which are bound together, in particular $\ell_I=+1$ ($\ell_I=-1$) for elementary qp (quasihole) excitations. Edge terms and kinetic energy corrections will be ignored in the Lagrangian \eqref{eq:LCS} in the following. The qps carry fractional charges $e^*_J/e = \sum_{I=1}^n t_I \l K^{-1} \r_{I J}$, see Ref.\cite{Wen1995}. Thus we note that the Chern number of the incompressible groundstate is given by the sum of the fractional charges of elementary topological excitations, $\C = \sum_J e^*_J / e$.

When the impurity (in a given internal state) binds $\ell_I$ qps of type $I$, their currents can be directly related to the TP current $j_\mu$ by $j_{I \mu} = \ell_I j_\mu$. This is demonstrated by a microscopic calculation in the main text. Using this expression and integrating out the auxiliary $U(1)$ gauge fields $a_{I \mu}$ in the Lagrangian \eqref{eq:LCS}, see Ref.\cite{Moore2014}, we obtain
\begin{multline}
\L_\eff = - \frac{e^2}{4 \pi}  ~ \uline{t}^T \uuline{K}^{-1} \uline{t} ~ \epsilon^{\mu \nu \lambda} A_\mu \partial_\nu A_\lambda +  \pi \uline{\ell}^T \uuline{K}^{-1} \uline{\ell}~  j_\mu \frac{\epsilon^{\mu \nu \lambda} \partial_\nu}{\partial^2} j_\lambda  \\
+ \l q B_\mu + e A_\mu ~ \uline{t}^T \uuline{K}^{-1} \uline{\ell} \r j_\mu,
\label{eq:LeffTPfull}
\end{multline}
where $\partial^2 = \partial_\tau \partial_\tau$. We also included a coupling $q B_\mu j_\mu$ of the impurity to an additional external field $B_\mu$. The first term in Eq. \eqref{eq:LeffTPfull} is a Chern Simons term for the external gauge field $A_\mu$. The second term describes the braiding statistics of the TP, which coincides with the expected qp statistics, see e.g. Ref.\cite{Altland2010}. The statistical phase $e^{i \theta}$ picked up when interchanging two TPs adiabatically is given by $\theta = \pi  \uline{\ell}^T \uuline{K}^{-1} \uline{\ell}$ \cite{Wen1995}. The last term, most important to our discussion, corresponds to an effective gauge field $\l A_\mu e^*_\TP + B_\mu q \r /e$ seen by the TP, where the $A_\mu$-charge of the TP is given by $e^*_\TP = e ~ \uline{t}^T \uuline{K}^{-1} \uline{\ell}$.

By applying the interferometric protocol introduced in the main text to different flavors $J=1,...,n$ of TPs, all the Chern numbers (or, equivalently, all the fractional charges)
\begin{equation}
\C_\TP^{(J)} = \frac{1}{\C_J} = \frac{e}{e^*_J}
\label{eq:CTPresult}
\end{equation}
of elementary topological excitations can be measured. To this end a single qp (or quasihole) of flavor $J$ is bound to the impurity. The last equation is then derived as Eq.\eqref{eq:CTPderivation} in the main text. When the TP Chern numbers of all qps are known, the Chern number of the incompressible groundstate can be derived from Eq.\eqref{eq:manyBodyChernKtheory},
\begin{equation}
\C = \sum_{J=1}^n \frac{(\pm 1)}{\C_\TP^{(J)}}.
\label{eq:CgsResult}
\end{equation}
Here $(-1)$ needs to be inserted if the elementary quasihole excitation of flavor $J$ is used, with $\ell_I = - \delta_{I,J}$, and $(+1)$ for elementary qps with $\ell_I = \delta_{I,J}$.

In the main text we discuss the case of $\nu=1/m$ Laughlin states where $n=t=1$ and $K=m$. In this case there exists one quasihole branch, with fractional charge $e^*=- e/m$. According to Eq.\eqref{eq:CTPresult} the Chern number of the TP consisting of an impurity bound to a hole is given by $\C_\TP = -m$. We confirm this by a microscopic calculation for a fractional Chern insulator in FIG.\ref{fig:FCIresults}. Then Eq.\eqref{eq:CgsResult} predicts a fractional Chern number $\C=1/m$ of the incompressible Laughlin state, in agreement with the established result by Niu et al.\cite{Niu1985}.\\

\noindent
\textbf{Exact polaron transformation}
To calculate the topological invariants characterizing TPs exactly, we need its full wavefunction $\ket{\psi_\TP(\vec{q})}$ for any given value of the total TP quasi-momentum $\vec{q}$. Here we develop a method allowing to calculate $\ket{\psi_\TP(\vec{q})}$ using exact numerical methods. Our approach is based on the Lee-Low-Pines (LLP) unitary transformation \cite{Lee1953} introduced in the context of conventional polaron physics, which makes the conservation of the polaron momentum explicit. The effect of the external force $\vec{F}$ acting on the impurity is also discussed in this framework. 

Our starting point is the impurity-centered LLP transformation
\begin{equation}
\hat{U}_{\text{LLP}}(t) = \exp \left[ i \hat{\vec{R}}_\I \cdot \l \hat{\vec{p}}_{\rm c} + \vec{F} t \r \right].
\label{eq:ULLPimpCentDef}
\end{equation}
To define the impurity position operator $\hat{\vec{R}}_{\I}$ and the fermion momentum operator $\hat{\vec{p}}_{\rm c}$, a gauge choice is made. We introduce the magnetic unit cell of size $a_x \times a_y$ for the fermion Hamiltonian $\H_0$, see Eq.\eqref{eq:H0fermiHofstadter}, which contains an integer number of flux quanta. Using the Landau gauge as in Eq.\eqref{eq:H0fermiHofstadter} and assuming $\alpha=r/s$ with $r,s$ integers, we have $a_x=a$ and $a_y=s a$. Next we label sites within the unit cell by an integer $\mu$ and define the impurity position operator
\begin{equation}
\hat{\vec{R}}_\I = \sum_{j_x,j_y,\mu} \underbrace{(j_x a_x, j_y a_y)^T}_{= \vec{r}_{\vec{j}}} ~ \bd_{\vec{j},\mu} \b_{\vec{j},\mu}.
\label{eq:RIDef}
\end{equation}
Here the integers $j_{x,y}$ label unit cells and $(\vec{j},\mu)$ is merely an alternative way of parametrizing the site indices $(m,n)$ which were used previously in the definition of the model. Hence we see that $\hat{\vec{R}}_\I$ represents only the position of the unit cell, but not the positions of individual sites \emph{within} one cell. Similarly the fermion momentum operator is
\begin{equation}
\hat{\vec{p}}_{\rm c} = \sum_{\vec{k},\mu} \vec{k}  ~ \cd_{\vec{k},\mu} \c_{\vec{k},\mu},
\end{equation}
where we introduced operators in momentum space $\c_{\vec{k},\mu} :=  \sqrt{\frac{a_x a_y}{L_x L_y}} \sum_{\vec{j}} e^{i \vec{k} \cdot (j_x a_x \vec{e}_x + j_y a_y \vec{e}_y)} \c_{\vec{j},\mu}$. The wave vector $\vec{k}$ takes quantized values $\vec{k} = 2 \pi \l i_x / L_x , i_y / L_y \r^T$ for integers $i_x=1,...,L_x/a_x$ and $i_y=1,....,L_y/a_y$ and with $L_{x,y}$ denoting system size in $x$ and $y$ direction. Note that although the impurity lattice has a smaller period of $a$ we have chosen the larger magnetic unit cell of the fermion model in Eq.\eqref{eq:RIDef}. This is necessary to distinguish between inequivalent sites $\mu$ within one magnetic unit cell for both the fermions and the impurity. 

We proceed by applying the LLP transformation defined above to the Hamiltonian $\H = \H_0 + \H_\I + \H_{\rm int}$, see Eqs.\eqref{eq:H0fermiHofstadter}, \eqref{eq:HimpTPhofstadterICI}. The new effective Hamiltonian in the polaron frame reads $\tilde{\mathcal{H}}(t) = \hat{U}^\dagger_{\text{LLP}}(t) \H \hat{U}_{\text{LLP}}(t) - i \hat{U}_\text{LLP}^\dagger(t) \partial_t \hat{U}_\text{LLP} (t)$. First we note that the many-body (fermion) Hamiltonian $\H_0$ trivially commutes with the LLP transformation because of its translational invariance by multiples of one magnetic unit-cell. The potential term $\vec{F} \cdot \sum_{m,n} \vec{r}_{m,n} \bd_{m,n} \b_{m,n}$ in Eq.\eqref{eq:HimpTPhofstadterICI} also commutes with the LLP transformation, as can easily be checked.

To transform the kinetic energy of the impurity, we introduce the single-particle band Hamiltonian $h^\I_{\mu ,\mu'}(\vec{k})$ defined in the magnetic unit-cell of the fermions. This allows us to write $\H_\I = \sum_{\vec{k}, \mu,\mu'} \bd_{\vec{k},\mu} \b_{\vec{k},\mu'} ~ h^\I_{\mu ,\mu'}(\vec{k})$ in the absence of the force $\vec{F}$. The momentum operators $\b_{\vec{k},\mu}$ are defined as in the case of fermions $\c_{\vec{k},\mu}$ discussed above. The impurity position operator $\hat{\vec{R}}_\I$ is the infinitesimal generator of displacements in quasi-momentum space. Therefore it holds
\begin{equation}
\hat{U}_\text{LLP}^\dagger(t)  \b_{\vec{k},\mu}  \hat{U}_\text{LLP}(t) = \b_{\vec{k} + \hat{\vec{p}}_{\rm c} + \vec{F} t,\mu}
\end{equation}
as can also be checked directly from the definition of the impurity momentum operators $\b_{\vec{k},\mu}$. Hence after the LLP transformation we obtain the effective impurity Hamiltonian
\begin{multline}
\tilde{\mathcal{H}}_\I(t) = \hat{U}_\text{LLP}^\dagger(t) \H_\I \hat{U}_\text{LLP}(t) - i \hat{U}_\text{LLP}^\dagger(t) \partial_t \hat{U}_\text{LLP} (t)  =\\
 = \sum_{\vec{k},\mu ,\mu'} h^\I_{\mu,\mu'}(\vec{k} - \hat{\vec{p}}_{\rm c} - \vec{F} t) ~ \bd_{\vec{k},\mu} \b_{\vec{k},\mu'}.
\label{eq:apdxH0LLP}
\end{multline}

Finally we apply the LLP transformation to the impurity-fermion interaction $\H_{\rm int}$. To keep the discussion general we consider a density-density interaction of the form
\begin{equation}
\H_\text{int} = \sum_{\vec{j}, \vec{i}, \mu, \nu} V_{\mu,\nu}(\vec{r}_{\vec{j}} - \vec{r}_{\vec{i}}) ~ \bd_{\vec{i},\mu}  \b_{\vec{i},\mu} \cd_{\vec{j},\nu}  \c_{\vec{j},\nu},
\end{equation}
where $V_{\mu,\nu}(\vec{r})$ denotes an arbitrary potential. Because the LLP transformation displaces the many-body system (recall that $\hat{\vec{p}}_{\rm c}$ is the infinitesimal generator of fermion translations), it holds 
\begin{equation}
\hat{U}_\text{LLP}^\dagger(t)  \c_{\vec{j},\mu}  \hat{U}_\text{LLP}(t) = \c_{\vec{j} - \hat{\vec{R}}_{\I},\mu}.
\end{equation}
Using this relation and restricting ourselves to the subspace of one impurity, we obtain
\begin{equation}
\hat{U}_\text{LLP}^\dagger(t)  \H_{\rm int} \hat{U}_\text{LLP}(t) = \sum_{\vec{i}, \mu} \bd_{\vec{i},\mu}  \b_{\vec{i},\mu}  ~  \sum_{\vec{j}, \nu}  V_{\mu,\nu}(\vec{r}_{\vec{j}} ) ~ \cd_{\vec{j},\nu}  \c_{\vec{j},\nu}.
\end{equation}
This corresponds to a static potential for the many-body fermion system, centered around the central unit-cell (where $\vec{r}_{\vec{j}}=0$). For the local interaction assumed in the main text it holds $V_{\mu,\nu}(\vec{r}) = V \delta_{\vec{r},\vec{0}} \delta_{\mu,\nu}$.

For a single impurity, the resulting Hamiltonian in the polaron frame (written in second quantization for notational convenience) is
\begin{multline}
\tilde{\mathcal{H}}(t) = \sum_{\vec{k},\mu,\mu'} \bd_{\vec{k},\mu}  \b_{\vec{k},\mu'} \otimes \Biggl[  h^\I_{\mu,\mu'}(\vec{k} - \hat{\vec{p}}_{\rm c} - \vec{F} t) +\\
+ \delta_{\mu,\mu'} \sum_{\vec{j},\nu} V_{\mu,\nu}(\vec{r}_{\vec{j}} ) ~ \cd_{\vec{j},\nu}  \c_{\vec{j},\nu}  \Biggr] + \H_0.
\label{eq:HeffTPexplctImpCent}
\end{multline}
Here we used the relation $\sum_{\vec{i}}  \bd_{\vec{i},\nu}  \b_{\vec{i},\nu} = \sum_{\vec{k}}  \bd_{\vec{k},\nu}  \b_{\vec{k},\nu}$ to make the conservation of the TP momentum apparent, $ [ \tilde{\mathcal{H}} , \sum_{\mu} \cd_{\vec{k},\mu} \c_{\vec{k},\mu} ] = 0$. Eq.\eqref{eq:HeffTPexplctImpCent} demonstrates that the force $\vec{F}$ directly couples to the TP momentum. Using exact diagonalization techniques we solved the TP band Hamiltonian \eqref{eq:HeffTPexplctImpCent} for different values of $\vec{k}$, see FIGs. \ref{fig:ICIresults} and \ref{fig:FCIresults}. This allows us to extract the TP Chern number $\C$ for systems with a small number of fermions.\\

%%%%%%%%%%%%%%%%%%%%%%%%%%%%%%%%%%%%%%%%%%%%%%%%%%%%%%%
\textbf{Acknowledgements}\\
%%%%%%%%%%%%%%%%%%%%%%%%%%%%%%%%%%%%%%%%%%%%%%%%%%%%%%%
We acknowledge useful discussions with M. Knap, M. Hafezi, F. Letscher, W. Phillips, J. Sau, J. Taylor, I. Bloch, E. Mueller, M. Greiner, F. Gerbier, J. Dalibard and C. Weitenberg. FG  acknowledges support by the Excellence Initiative (DFG/GSC 266), the ``Marion K\"oser Stiftung" and the Moore foundation. ED and FG acknowledge support from Harvard-MIT CUA, NSF Grant No. DMR-1308435, AFOSR Quantum Simulation MURI, the ARO-MURI on Atomtronics and the ARO MURI Quism program. ED acknowledges support from the Humboldt foundation,  Dr.~Max R\"ossler, the Walter Haefner Foundation and the ETH Foundation. DA acknowledges support by the Alfred Sloan Foundation. NYY  acknowledges support from the Miller Institute for Basic Research in Science.\\

%%%%%%%%%%%%%%%%%%%%%%%%%%%%%%%%%%%%%%%%%%%%%%%%%%%%%%%%
%\textbf{Author contributions}\\
%%%%%%%%%%%%%%%%%%%%%%%%%%%%%%%%%%%%%%%%%%%%%%%%%%%%%%%%
%All authors contributed substantially to the writing of the manuscript. FG performed the calculations. FG and ED conceived the method. NYY, DA and MF contributed to clarifying the initial idea of the method.\\
%
%%%%%%%%%%%%%%%%%%%%%%%%%%%%%%%%%%%%%%%%%%%%%%%%%%%%%%%%
%\textbf{Competing financial interests}\\
%%%%%%%%%%%%%%%%%%%%%%%%%%%%%%%%%%%%%%%%%%%%%%%%%%%%%%%%
%The authors declare no competing financial interests.

\def\bibsection{\section*{\refname}} 

%\bibliography{/Users/fgrusdt/Documents/Library/dataBase_JabRef2.bib}
%\bibliography{/Users/fgrusdt/Documents/PhD/JabRef/dataBase_JabRef2.bib}


\newpage

\onecolumngrid

~\\

\begin{center}
\large \textbf{Supplementary Material: Interferometric Measurement of Many-body \\ Topological Invariants using Mobile Impurities}
\end{center}

~\\

In this supplementary we discuss the case when the quasiparticle charge $e^*/e=p/q$ is not the inverse of an integer, $p \neq 1$, and $p$, $q$ are relative prime. In this case, the topological field theory of the topological polaron (TP) presented in the main text predicts a Chern number 
\begin{equation}
\C_\TP=\frac{q}{p}.
\end{equation}
Because the TP Chern number needs to be an integer due to gauge invariance, the groundstate manifold has to be at least $p$-fold degenerate. In that case all states share a total Chern number $\C_\TP^{\rm tot}=q$.

Here we explain the origin of this degeneracy (incommensurability of unit cells) and investigate numerically the $\nu=1/3$ Laughlin state. To this end we bind $p=2$ quasiholes (qhs) to the impurity, which realizes the simplest non-trivial fraction $e^*/e=-2/3$. Indeed we find a robust two-fold groundstate degeneracy and the Berry curvature is consistent with predictions from the topological field theory.

\subsection{Topological field theory}
To understand the origin of the two-fold groundstate degeneracy of the TP, let us recall our argument from the main text. The impurity is bound to two fractionally charged qhs with $e^*/e=-1/3$ each. Every qh sees the original host particles as a source of magnetic flux. In a mean-field theory this leads to a homogeneous effective magnetic field $b_z^*=-\frac{2}{3} b_z$ seen by the impurity. 

The reasoning above relies on the effective low-energy topological field theory, which does not take into account lattice effects. Let us now treat the lattice as a perturbation, which is justified when the magnetic flux per plaquette $\alpha \ll 1$ is sufficiently small. The lattice defines the magnetic unit cell of the TP, see Methods Section of the main text. Its size is equal to that of the original magnetic unit cell of the host particles, $2 \pi \ell_{\rm B}^2$. 

\begin{figure}[b]
\centering
\epsfig{file=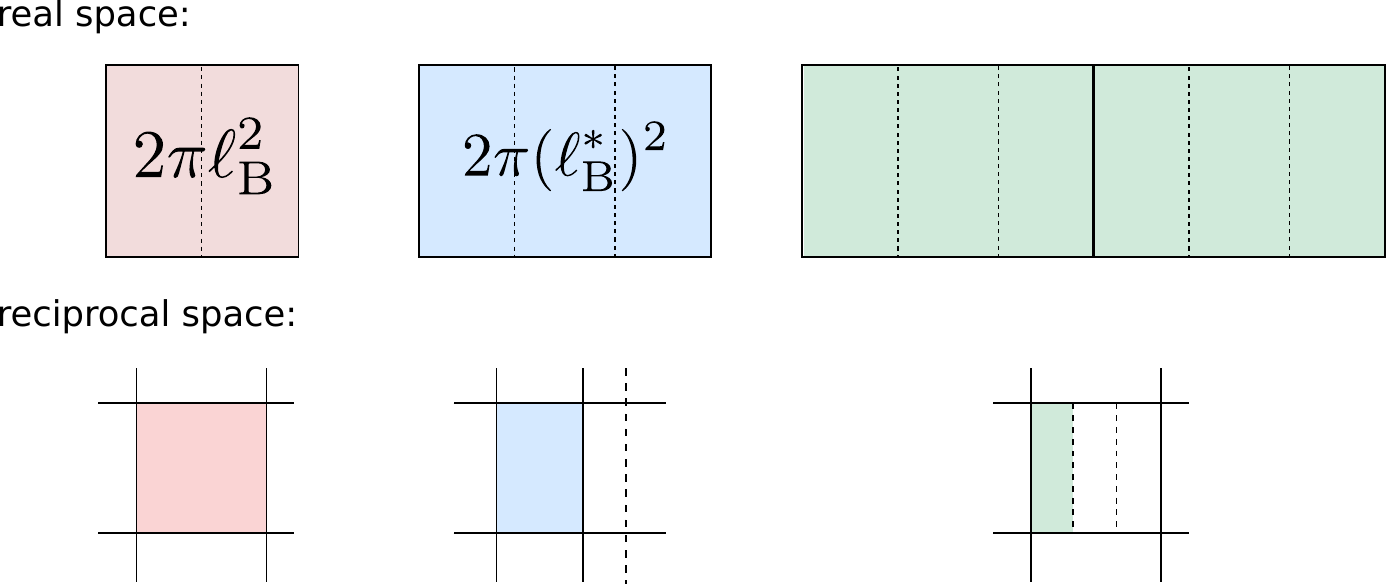, width=0.5\textwidth}
\caption{The size of the magnetic unit cell of host atoms is incommensurable with the size of the effective magnetic unit-cell seen by the impurity bound to two fractionally charged quasiparticles. The smallest commensurable unit cell is shown in on the right. In the second row the corresponding magnetic BZs are sketched.}
\label{fig:TPdegeneracy}
\end{figure}

The size of the magnetic unit cell in the effective field $b_z^*$, on the other hand, is larger by a factor of $3/2$, $2 \pi \ell_{\rm B}^{*2} = 3 \pi \ell_{\rm B}^2$, see FIG.\ref{fig:TPdegeneracy}. To construct groundstates in the full lattice model from the groundstate in the effective continuum model, we choose the smallest commensurable unit cell, with a size $6 \pi \ell_{\rm B}^2$. 

In reciprocal space, we thus arrive at a description of the system in a reduced zone with size $1/3$ of the original magnetic Brillouin zone (BZ). Because it has half the size of the effective magnetic BZ, it contains two degenerate groundstates. 

In this way we constructed two-fold degenerate groundstates in the original BZ of host particles. By taking into account lattice effects, small gaps may open which can be different everywhere in the BZ. Thus we understand the two-fold degeneracy as a consequence of the incommensurability of the original magnetic unit cell and the effective magnetic unit cell seen by the impurity.

%%%%%%%%%%%%%%%%%%%%%%%%%%%%%%%%%%%%%%%%%%%%%%%%%%%%%
%%%%%%%%%%%%%%%%%%%%%%%%%%%%%%%%%%%%%%%%%%%%%%%%%%%%%

\subsection{Numerical results}
To check our prediction of a two-fold degenerate ground state, we simulated a fractional Chern insulator at filing $\nu=1/3$ with $N=2$ fermions and with two qh excitations. To make our numerics more robust, we implemented the Kapit-Mueller lattice model \cite{Kapit2010}, where, instead of the nearest neighbor hoppings from Eq.(9) in the main text, long range tunnelings as suggested in Ref.\cite{Kapit2010} are used. This leads to flat bands, reducing the quasiparticle dispersion in the TP Hamiltonian. The impurity Hamiltonian, see Eq.(10) in the main text, is unchanged. To trap two quasiholes efficiently, we added additional nearest neighbor interactions of strength $V/2$ to the local interaction $\H_{\rm int}$ below Eq.(10) in the main text.

The resulting TP bandstructure has a numerically exact two-fold groundstate degeneracy. Also the higher bands are two-fold degenerate. This is in agreement with our theoretical prediction. 

%%%%%%%%%%%%%%%%%%%%%%%%%%%%%%%%%%%%%%%%%%%%%%%%%%%%%
\begin{figure}[t]
\centering
\epsfig{file=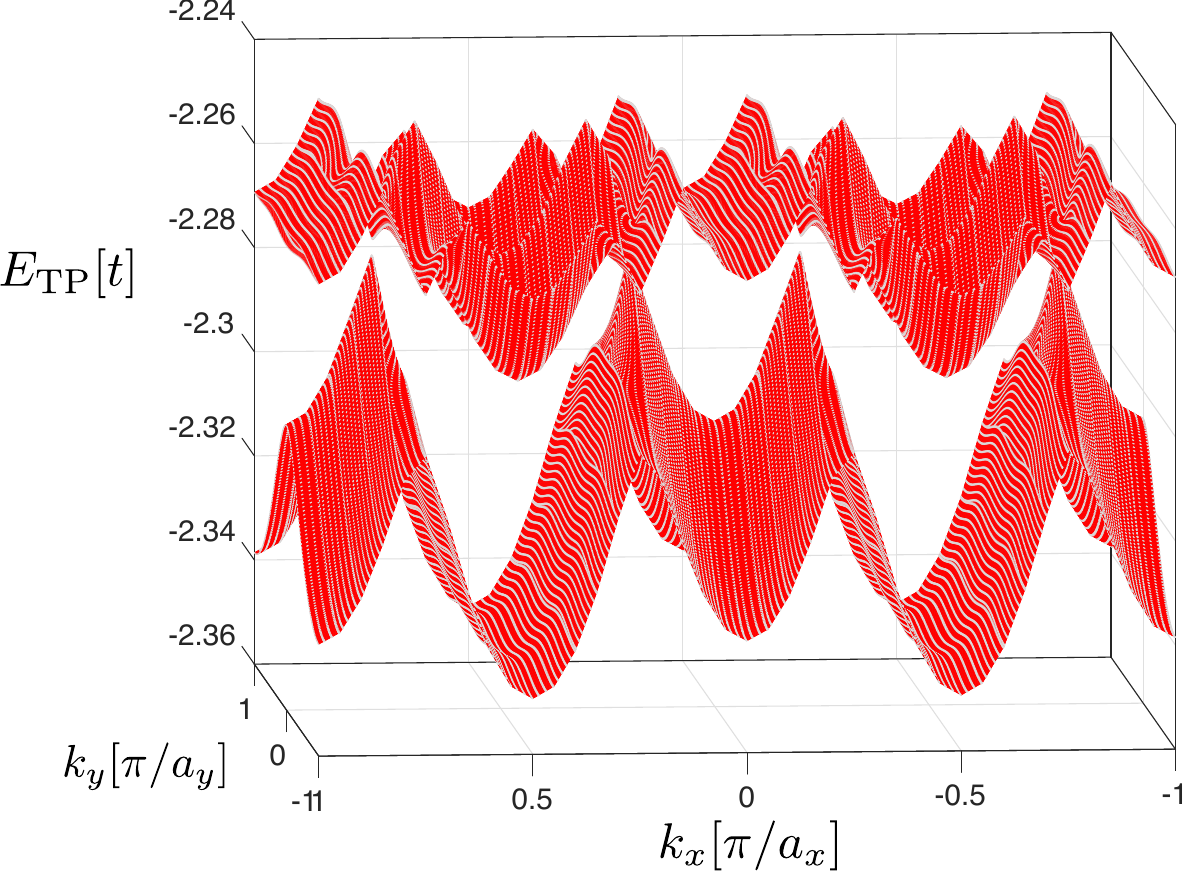, width=0.48\textwidth}
\caption{The lowest four bands of the TP bandstructure are shown. The bands come in degenerate pairs. The model is described in the text.}
\label{fig:TPBandStructure}
\end{figure}
%%%%%%%%%%%%%%%%%%%%%%%%%%%%%%%%%%%%%%%%%%%%%%%%%%%%%

%%%%%%%%%%%%%%%%%%%%%%%%%%%%%%%%%%%%%%%%%%%%%%%%%%%%%
\begin{figure}[b]
\centering
\epsfig{file=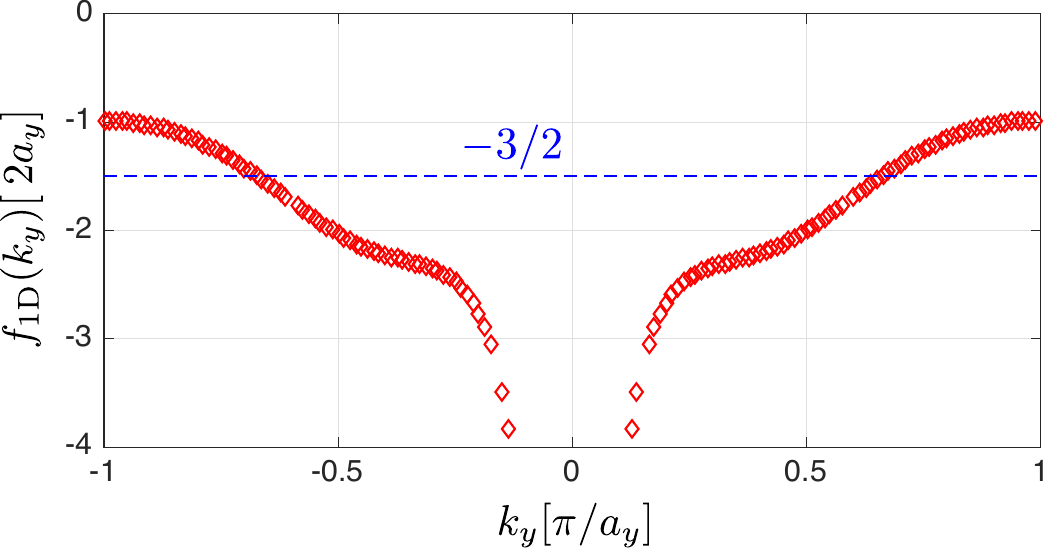, width=0.4\textwidth}
\caption{The integrated Berry curvature $f_{\rm 1D}(k_y)$ of the two degenerate bands, see Eq.\eqref{eq:f1Ddef}, is shown. It is compared to the constant value $-3/2$ expected from the low-energy field theory.}
\label{fig:TPBerryCurvature}
\end{figure}
%%%%%%%%%%%%%%%%%%%%%%%%%%%%%%%%%%%%%%%%%%%%%%%%%%%%%

The lowest bands of the TP bandstructure are shown in FIG.\ref{fig:TPBandStructure}. We simulated $N=2$ fermions on a $8 \times 4$ lattice with $\alpha=1/4$. This corresponds to $N_\phi=8$ flux quanta in the system. The impurity hopping was $J=0.1 t$ and the interaction strenghts were $U=10 t$ and $V=0.5 t$.

In FIG.\ref{fig:TPBandStructure} we observe additional degeneracies between the bands, found on the axis $k_y=0$. This suggests that the microscopic binding of the quasiholes to the impurity breaks down at these values of TP momenta. We think that this is a finite size effect, and the microscopic binding of qhs to the impurity deserves a more careful analysis which will be devoted to future work. 

Finally, we calculated the Berry curvature ${\rm tr} \mathcal{F}(\vec{k})$ of the TP, where the trace is over the two degenerate states. In FIG. \ref{fig:TPBerryCurvature} we plot the integrated curvature along $x$-direction,
\begin{equation}
f_{\rm 1D}(k_y) = \int_{-\pi/a_x}^{\pi/a_x} dk_x  {\rm tr} \mathcal{F}(\vec{k}).
\label{eq:f1Ddef}
\end{equation}
We compare it to the value $f_{\rm 1D}^0 = -3 \times 2 a_y$ expected for a homogeneous Berry curvature which gives rise to the expected Chern number $\C_\TP^{\rm tot}=-3$. Away from $k_y=0$ the average Berry curvature agrees with the predicted value $f_{\rm 1D}^0$. Around $k_y=0$, on the other hand, the Berry curvature becomes large, which we attribute to an effect of the additional degeneracies with higher bands identified in FIG.\ref{fig:TPBandStructure}.

\def\bibsection{\section*{\refname}} 

%\bibliography{/Users/fgrusdt/Documents/Library/dataBase_JabRef2.bib}
%\bibliography{/Users/fgrusdt/Documents/PhD/JabRef/dataBase_JabRef2.bib}


\end{document}